\documentclass[preprint,proceedings]{rmaa}
\usepackage{paralist}
\usepackage{psfrag,color}

\SetYear{2004}
\SetConfTitle{Compact Binaries in the Galaxy and Beyond}

\title{Some updates on the role of Magnetic Fields in Cataclysmic Variables}

\author{Coel Hellier\altaffilmark{1}}
\altaffiltext{1}{Keele University, UK.}

\shortauthor{Hellier} 
\shorttitle{Magnetic Fields in Cataclysmic Variables}

\fulladdresses{
\item Coel Hellier: Keele University, Staffordshire, ST5 5BG, U.K.
  (\email{ch@astro.keele.ac.uk}).} % Note final period. 

\listofauthors{C. Hellier}
\indexauthor{Hellier, C.}

\abstract{In this review talk I cover some recent developments
in understanding the role that magnetic fields play in cataclysmic
variables. I discuss the recent DNO--QPO unification models; the
disk--magnetosphere boundary; some issues concerning the soft
blackbody component and the nature of the X-ray spectra in MCVs;
whether the SW Sex stars are magnetic, and finally I mention the 
weird behavior of FS~Aur and HS\,2331+3905.}

\resumen{Spanish abstract -- to be supplied by the editors.}

\addkeyword{Accretion, accretion disks}
\addkeyword{Novae, cataclysmic variables}

\begin{document}
\maketitle

\section{DNO--QPO unification}
A major advance of recent years is a new understanding of the
dwarf-nova oscillations (DNOs) and quasi-periodic oscillations (QPOs)
seen in dwarf novae in outburst.  Such things have been
recorded and studied since the 1970s, but only recently, with a series
of papers by Warner and Woudt, do we have a compelling account of their
origin (Woudt \&\ Warner 2002; Warner \&\ Woudt 2002; Warner, Woudt \&\
Pretorius 2003).

The heart of the model is a suggestion that, during DN outbursts, an 
equatorial belt of the white dwarf is spun up by enhanced accretion. 
The belt sliding over the white-dwarf core results in a dynamo,
amplifying a seed field to the point where the field controls the
accretion flow near the white dwarf by carving out a magnetosphere
(Figure 1). 

The standard DNOs are simply pulsations at the rotation period of this
magnetosphere.  The transience of the magnetosphere explains why the
DNOs aren't seen in quiescence, while the low moment of inertia of the
belt explains the low coherence of the oscillations --- both of which
had previously been problems for a magnetic explanation of DNOs.

The second element of the model is the idea that the magnetic
field, playing on the inner edge of the disc, excites slow-moving
waves which run prograde round the disk with a period $\approx$\,15
times the magnetospheric spin period.  These bulges modulate the
light by simple obscuration, resulting in QPOs with a characteristic
$P_{\rm QPO} \approx 15 P_{\rm DNO}$.  

A second type of DNO then results from reprocessing of the first DNO
off the QPO bulges, giving the beat relation $1/P_{\rm DNO2} =
1/P_{\rm DNO1} - 1/P_{\rm QPO}$.  Further, Warner \etal\ (2003) claim
a third type of DNO, which they suggest results from the rotation
of a field attached to the body of the white dwarf, not the
spun-up belt. This is rotating more slowly and so produces `long
period DNOs'.  All of this interpretation is supported by an impressive
amount of observational documentation.

\begin{figure*}[!t]
\hspace*{0.15\textwidth}\includegraphics[width=0.7\textwidth]{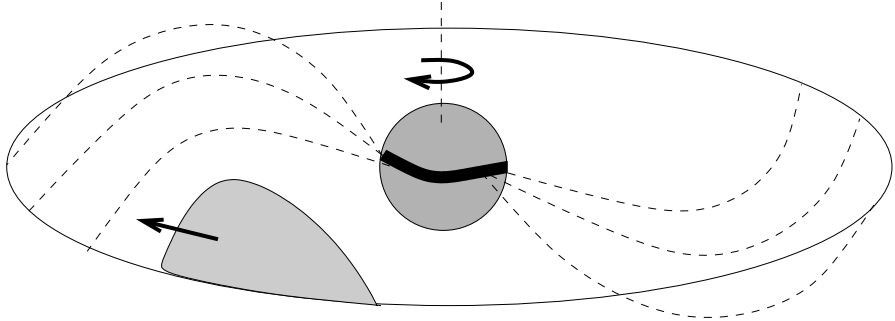}%
  \caption{A schematic of Warner \&\ Woudt's DNO/QPO model, with dynamo
action in a spun-up belt producing a transient magnetosphere, which excites
travelling waves on the inner edge of the disk.}
\end{figure*}

The above theory raises the question of whether the same QPO bulges
are excited at the inner edge of disks in intermediate polars.
Certainly, such bulges are the most plausible explanation for the
5000-s QPOs seen in GK~Per during outburst (e.g.\ Hellier, Harmer \&\
Beardmore 2004), but such things are not generally reported in IPs in
quiescence.  One reason might be observational: with a typical 1000-s
spin period, the QPO period would be $\sim$\,4 hrs, and it is hard to
observe for the dozen cycles that would be needed to prove the
presence of a low-level, incoherent modulation. Further, such signals
might be masked by orbital-cycle variations. 

A second explanation
might be that the bulges are only excited when there is strong
slippage between the magnetic field and the inner disk, and that this
isn't so in intermediate polars in their equilibrium, quiescent state
--- a topic that seems worth pursuing observationally.

An interesting point (noticed by Warner \&\ Woudt 2002 and
independently by Mauche 2002) is the fact that the QPO/DNO ratio can
be extended over 5 orders of magnitude to cover the much faster QPOs
seen in neutron-star and black-hole binaries.  Is this coincidence or
does it imply a causal similarity?  There is valid skepticism over any
link, since of course black-holes don't have a surface, nor a
permanent magnetic field, and thus are unlikely counterparts of white
dwarfs.  However, in the model above the field is also a transient one
created by a dynamo, and the action occurs at the interaction of the
field with the inner disk, with the solid surface playing little
role. Thus the situations are not as dissimilar as they may at first
appear.

\section{The disk--magnetosphere boundary}
While on the subject of the disk--magnetosphere boundary, it bears
restating that this is one of the least understood regions of a CV.  A
paper on FO~Aqr by Evans \etal\ (2004) shows 
that the accretion curtain appears to be swept back, trailing the
magnetic pole by a quarter of a cycle.  The opposite was found in PQ~Gem
(Mason 1997), where the accreting field lines lead the pole.
One can then ask whether these twists are related to disk--field
disequilibrium and thus to the torques on the white dwarf.  At first
sight it appears so, since the white dwarf in PQ~Gem is spinning down
whereas that in FO~Aqr is currently spinning up.  However, FO~Aqr has
changed from a period of spin-down to one of spin-up, without any
obviously related change in the spin-pulse profile. Thus any
interpretation is problematic, and the whole issue of the disk--field
interaction and the resulting torques is one that could do with 
more study.

\section{The footprint}
Turning now to the accretion footprint on the white dwarf, Ramsay \&\
Cropper (2004) have proposed a major re-evaluation of the accretion
process in AM~Her stars.  For two decades it has been conventional
wisdom that the AM~Her stars show a strong `soft excess' over that
expected in the simplest accretion model (a hard-X-ray-emitting shock 
which irradiates the white-dwarf surface, resulting in soft blackbody 
emission that amounts to half the total flux).   The excess is usually
attributed to `blobby accretion' in which blobs of material do not
shock, but penetrate the white-dwarf surface and thermalize, greatly
boosting the soft/hard ratio. 

Now, from a systematic analysis of {\sl XMM\/} data, Ramsay \& Cropper
(2004) find that most AM~Her spectra are indeed compatible with the
simple model, and that previous reports of soft excesses were, to a
large extent, artefacts of calibration and band-pass uncertainties
(Figure 2).

However, a small number of systems do show a large soft excess, 
and are presumably dominated by blobby accretion.  But why?  Ramsay
\&\ Cropper discuss the obvious variables such as field strength, but
find no obvious correlation with the presence of a soft excess. 

A similar question arises in the intermediate polars.  Since {\sl
Rosat\/} (e.g.\ Haberl \&\ Motch 1995) we've known of a minority of
IPs that show a soft-blackbody component, but the majority do not.  
Again, we have no good explanation for the difference, and no obvious
correlation with field strength or other variables to guide us.

V405~Aur is one of the IPs with soft blackbody emission, and is also
peculiar in that it shows a single-humped spin pulse at hard-X-ray 
energies but a double-humped pulse at softer energies.  One idea
explains the difference between single-humped and double-humped 
IPs as an absorption effect: IPs with short, fat accretion columns 
beam X-rays upwards and so produce double-humped pulsations, whereas IPs 
with tall, thin columns beam X-rays sideways and produce  
single-humped pulsations (e.g.\ Hellier 1995; Allan \etal\ 1996). 

However, an analysis of {\sl XMM\/} data by Evans \&\ Hellier (2004)
shows that, at least in V405~Aur, the double-humped soft pulse
is not the product of absorption. Instead it is a modulation of the
visible area of the blackbody component, resulting primarily from 
variable foreshortening of the heated polecaps as the
white dwarf rotates.   It remains to be seen what this implies 
for the double-humped optical pulsation, and whether the findings are
applicable to other double-humped IPs.

\begin{figure}[!t]
\includegraphics[width=\columnwidth]{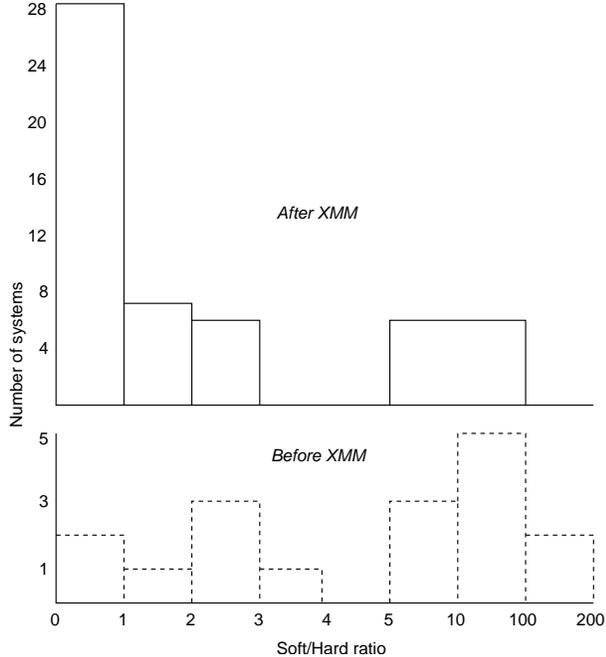}%
  \caption{Ramsay and Cropper's reassessment of the energy balance in
AM~Her stars. Before {\sl XMM\/} (bottom) it was thought that most
AM~Hers showed a strong soft excess. Now (top) only a minority show a
soft excess, and the majority are compatible with the standard model.}
\end{figure}

\section{The X-ray spectra}
The standard model for X-ray spectra in IPs invokes a stratified
column in which material cools beneath the accretion shock.
By assuming optically thin, collisionally ionized
emission, one can use a code such as {\sc mekal} and sum the emission
between the hot shock and the point where the column becomes optically
thick as it merges with the white dwarf.    Such a model gives an
excellent representation of the spectrum of EX~Hya (Cropper \etal\ 2002).

However, Mukai \etal\ (2003) report that EX~Hya is unusual.  From an
analysis of {\sl Chandra\/} grating spectra they find that most IPs
are not compatible with the above model. Instead, they obtain a better
fit with a photoionization code. This raises the issue of where the
emission arises.  One possibility is that the X-ray lines are
predominantly from photoionized pre-shock material. However, as
discussed by Hellier \& Mukai (2004), Doppler shifts of the lines are
of order $\sim$\,100 km s$^{-1}$, rather than the $\sim$\,1000 km
s$^{-1}$ expected for material approaching the shock at near the
escape velocity.  Such low velocities imply an origin near the base of
the accretion column where the material has been vastly decelerated
(which is also expected in the standard model, since these are the
densest regions and emission scales with density squared). But this
leaves us with no coherent model for the spectral characteristics
of the majority of IPs.

EX~Hya is atypical, possibly owing to it being below the period gap
and so having a much lower luminosity, which perhaps results in it
being easier to model.  Its importance will increase further now
that we know its distance to high precision, given Beuermann \etal 's
(2003) report of a parallax distance of 64.5\,$\pm$\,1.2 pc.

With EX~Hya's parameters now securely known, Beuermann \etal\ report
that the secondary is undermassive, being 18--30\%\ larger than a ZAMS
star.  Reassurringly, this corroborates the result of a large study of
superhumps by Patterson \etal\ (2003), which shows that secondary
stars in dwarf novae below the gap are 18\%\ larger than ZAMS.  From
the relatively small scatter in the values of superhump period excess,
Patterson \etal\ were also able to conclude that there is only one
evolutionary track leading to these stars, most likely without any 
nuclear evolution of the secondaries.

\section{Are the SW Sex stars magnetic?}
The SW~Sex phenomenon appears to be widespread in CVs ($>$\,20 systems
show at least some SW Sex characteristics) and is present in at least
one LMXB (Hynes \etal\ 2001).  It is thus important for our
understanding of accretion.

Many recent authors have favored models which invoke magnetic fields
to explain the SW~Sex characteristics, regarding SW~Sex stars as a
variant of the IPs (e.g.\ Groot \etal\ 2001; Rodr\'{\i}guez-Gil \etal\
2001; Hoard \etal\ 2003 and references therein; but see Hellier 2000
for a non-magnetic model).  Such models are supported by suggestions
of observed periodicities, including reports of periodic modulations
in polarization data that, if verified, would clinch the magnetic
nature of these stars.

However, at the risk of being thought unduely skeptical and ultimately
proved wrong, I note that no periodicity has yet been corroborated 
by multiple datasets or by independent groups --- something that usually
happens quickly for new IPs --- and that CVs are notorious for 
flickering behavior that can mimic periodicities in limited datasets. 

My main reason for skepticism about the magnetic nature of SW~Sex
stars is their general lack of X-ray emission, and particularly pulsed
X-rays.  IPs emit copious X-rays, with an obvious coherent pulsation.
Polarization is much harder to find, and pulsed polarization has been
seen in only a tenth of the known IPs.

In contrast, if the claims for SW~Sex stars are true, their fields are
strong enough to dominate the emission-line behavior, and to produce
phase-variable polarization, but we do not see pulsed X-rays. This
discrepancy, if true, would be telling us something fundamental
about accretion.
 
A further implication concerns VY~Scl stars.  There has
been a long-standing problem over the lack of dwarf-nova outbursts
in the low states of VY~Scl stars, given that disk-instability models
predict that they should occur.   One idea is that irradiation 
keeps the inner disk too hot for such outbursts (Leach \etal\ 1999), but 
Hameury \&\ Lasota (2002) prefer a model in which a strong magnetic
field evacuates the inner disk.  They note that many SW~Sex stars 
show VY~Scl low states, and cite the magnetic models as support of
their VY~Scl hypothesis.  

It is clear that the explanation of SW~Sex behavior has wide
implications.  It would thus be good to have corroboration of
periodicities sufficient to convince even an ardent skeptic, or to
have sufficent null results to settle the matter the other way.

\section{The weird stars FS~Aur and HS\,2331+3905}
As a last topic I turn to the stars FS~Aur and HS\,2331+3905, although
it is unclear whether the issue concerns magnetic fields or some sort 
of disk precession.  FS~Aur has an 86-min orbital period but also
shows a large-amplitude photometric modulation at 3.4 hrs 
(Tovmassian \etal\ 2003). No superhump has been seen with a period so much
longer than the orbital period, but the period is also too short to be 
disk precession. Could it be the spin period of a magnetic white
dwarf?  Well, so far we know of no system with a spin period longer
than the orbital period.

The issue becomes even stranger with the discovery of  HS\,2331+3905
(Araujo-Betancor \etal\ 2004). This star has an 81-min orbital period
and a 3.5-hr periodicity, making it similar to FS~Aur.
However, in FS~Aur the 3.4-hr periodicity appears in photometry
only, and not in radial velocities, whereas in HS\,2331+3905 the 3.5-hr 
periodicity is seen in radial velocities, but not in photometry.
Explaining 3-hr periodicities in 80-min binaries is hard enough 
without having to explain that also!

\end{document}